\documentclass[aps,prl,preprint,grouppedaddress,showpacs,showkeys]{revtex4}

\usepackage{graphicx}
\usepackage{dcolumn}
\usepackage{amsmath}
\usepackage{multirow}
\usepackage{epsfig}
\usepackage{amssymb}
\usepackage{color}

\begin{document}

\title{Effects of orbital selective dynamical correlation on the spin
susceptibility and superconducting symmetries in Sr$_2$RuO$_4$}

\author{Chang-Youn Moon}
\email{cymoon@kriss.re.kr}
\affiliation{Advanced Instrumentation Institute, Korea Research Institute of Standards and Science, 
Yuseong, Daejeon 305-340, Republic of Korea}

\date{\today}

\begin{abstract}
We investigate the connection between the local electron correlation and the momentum dependence
of the spin susceptibility and the superconducting gap functions in Sr$_2$RuO$_4$, using
density-functional theory combined with dynamical mean-field theory. Adopting frequency-dependent 
two-particle vertex moves the zero energy spin susceptibility peaks towards the Brillouin zone center, 
compared with
random-phase approximation which basically retains the peak positions closer to the Brillouin zone boundary
as determined by the Fermi-surface nesting. We find that $d_{xy}$ orbital plays a central role here via its 
enhanced correlation strength. Solving the linearized Eliashberg equation from this 
spin susceptibility, prime candidates of the superconducting gap symmetry are a $s$-wave, along with a nearly degenerate $d$-wave solution, all in spin singlet. Furthermore, another set of degenerate 
spin singlet gap functions emerges, odd with respect to $k$-point as well as orbital exchanges. We show 
that the stability of these gap functions are strongly dependent on the peak position of the spin 
susceptibility in the Brillouin zone.
\end{abstract}


\maketitle

Sr$_2$RuO$_4$ remains as one of most intriguing superconductors with the possibility of highly exotic 
and unconventional superconductivity, even after three decades since its discovery \cite{Maeno}.
Noticing the analogy with superfluid $^3$He, earlier studies kindled discussions on the 
$p_x+ip_y$ pairing \cite{MackenzieMaeno,RiceSigrist,Sigrist,MaenoKittaka,Kallin}. 
This order parameter can be characterized by two essentially independent properties; the chirality,
implying non-zero orbital magnetic moment, and the spin-triplet pairing. Spin-triplet pairing scenario, 
however, has been put into a serious question 
after recent spin susceptibility measurements using nuclear magnetic resonance (NMR) Knight shift
\cite{Pustogow,IshidaManago} and polarized neutron scattering \cite{Petsch}. They show that 
the susceptibility is suppressed below $T_c$ more consistently with the spin-singlet pairing, 
overruling an earlier contradicting Knight shift measurement \cite{IshidaMukuda}. The chirality, 
or time-reversal symmetry breaking (TRSB) more generally, has been supported by a recent zero-field 
muon spin relaxation (ZF-$\mu$SR) measurement under uniaxial stress resulting in split superconducting 
and TRSB transitions \cite{Grinenko}, together with earlier works using ZF-$\mu$SR \cite{Luke} 
and non-zero Kerr rotation \cite{Xia,Kapitulnik}. Evidences of the two-component order parameter 
\cite{Benhabib,Ghosh} also have appeared, which can serve as a broader
constraint on the nature of pairing in which the possibility of the chirality is included.
Combinations among even parity order parameters of such as $s$, $d$, and $g$ symmetries
are now generally considered plausible candidates.

The local electronic configuration near the Fermi energy ($E_F$) comprises four electrons 
in $t_{2g}$ levels split from empty $e_g$ levels within Ru $d$ states, and hence Sr$_2$RuO$_4$ 
is a multi-orbital system where Hund's coupling $J_H$ plays a significant role in local Coulomb 
correlations. Consequently its normal-state behaviors, such as the crossover from high-temperature 
incoherent phase to low-temperature Fermi-liquid \cite{MackenzieMaeno,Katsufuji,Imai,Hussey,Valla,Wang,Kidd}
are understood in the context of Hund's metals \cite{MravljeAichhorn,Medici,MravljeGeorges,Kugler}. 
General features of Hund's metals include strong local spin fluctuation and orbital 
differentiation \cite{HauleNJP,Moon}, as manifested in Sr$_2$RuO$_4$ with much enhanced 
effective mass of $d_{xy}$-derived Fermi surface (FS) sheet compared with $d_{xz/yz}$ derived 
ones \cite{MackenzieMaeno,MravljeAichhorn}. Then it is an important question that how the local 
correlation that governs the normal state works on the superconductivity. Assuming
the spin-fluctuation-mediated superconductivity, the two-particle vertex, which forms fully interacting 
spin susceptibility ($\chi$) when combined with the polarization bubble ($\chi^0$), is one of major 
channels through which the local correlation effect is incorporated in electron pairing.
While some of previous theoretical studies adopted the random-phase-approximation (RPA) for 
the vertex keeping the lowest order scattering process \cite{RomerScherer,RomerKreisel,RomerHirschfeld,RomerMaier,Gingras1,Gingras2}, 
others found that using frequency dependent vertex produces qualitative differences in susceptibility
and superconductivity \cite{Strand,EPL,Kaser}. However, distinctive roles played by the orbital selectivity and
its dynamic nature in the susceptibility evaluation remain elusive especially with respect to the features 
critically connected to the pairing symmetry, such as the peak position of the susceptibility.

In this study, we demonstrate how the dynamic local correlations shape the susceptibility 
in the $k$-space and
the pairing symmetry in Sr$_2$RuO$_4$ within the framework of density-functional theory combined with dynamical 
mean-field theory (DFT+DMFT). Adopting the dynamic vertex results in the peak position of spin susceptibility
closer to the Brillouin zone (BZ) center compared with RPA case, in better agreement with
experimental measurements. We argue it originates from the stronger frequency dependence of
the vertex within the $d_{xy}$ orbital than the other $t_{2g}$ components, relocating the peak
off the position determined by simple FS nesting as in the RPA case. When the linearized-Eliashberg
equation is constructed from the susceptibility and solved, the most probable gap function turns out 
to be of a $s$-wave symmetry with nodes followed by a $d$-wave solution slightly less stable, suggesting a
possible two-component order parameter with an accidental degeneracy. Another plausible 
solutions are found to be symmetry-protected doubly degenerate ones which are odd with
respect to both $k$ and orbital exchanges. We show that the relative stability of these gap functions changes drastically between
the respective susceptibility peak positions from the dynamical vertex and RPA, emphasizing on
the importance of the accurate evaluation of the susceptibility in $k$-space and of the orbital 
selective and dynamic local correlation effects which enable it. 

We use the modern implementation of DFT+DMFT method within all electron 
embedded DMFT approach \cite{DMFT} which is based on Wien2k \cite{wien2k}, without downfolding or other approximations. The code is freely available on the web \cite{Haulecode}. 
We employ LDA exchange-correlation functional \cite{LDA1,LDA2}, and the quantum impurity
model is solved by the continuous time quantum Monte Carlo (CTQMC) impurity solver 
\cite{CTQMC}. Here sampled quantities are expanded in the basis function obtained
by the singular value decomposition of the kernel for analytic continuation, 
ensuring reduced high-frequency noise \cite{Shinaoka}. Internal atomic positions are optimized, and $U = 4.5$ eV and $J = 1.0$ eV 
are adopted consistently with a previous
study on this material \cite{DengHaule} employing the same computational
method.
We use the Slater
parametrization of the Coulomb interaction in this study, and our $U$ and $J$
parameters are defined with respect to the three Slater parameters in such a way that $F^0= U$, $F^2=112/13~J$,
and $F^4=70/13~J$. BZ integration is done on the 30$\times$30$\times$30
k-point mesh for the body-centered tetragonal primitive unitcell in
converging charge density and self energy, while 40$\times$40$\times$2 mesh in the BZ for the conventional tetragonal unitcell is used for the susceptibility and gap function calculations.
All calculations are done in 116 K. 
Spin-orbit coupling (SOC) can have
non-negligible effects on the electronic structure mainly near the region where
different sheets of FS intersect, by strong orbital mixing \cite{ZhangGorelov,Kim,Tamai}. 
As evaluating the two-particle vertex including SOC is still not available, and also 
considering susceptibilities are less affected by SOC than one-particle spectra \cite{Gingras1},
we neglect SOC in this work.

\begin{figure}[tp]
\includegraphics[width=0.6\linewidth]{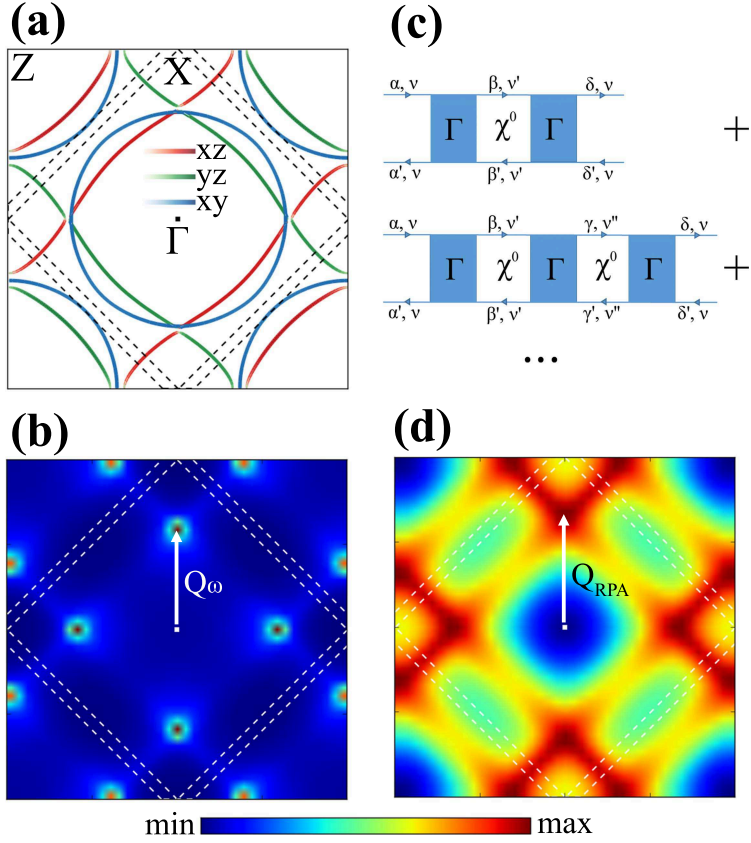}
\caption{(a) Fermi surface shown on the $k_z=0$ plane. Orbital character is 
denoted by the depth of the color as well as the thickness of the line.
(b) $[\Gamma\chi\Gamma]^s_{\alpha\alpha;\alpha\alpha}(i\nu,i\nu)$ 
where $\alpha=d_{xy}$ and $i\nu\approx 0$ on the $k_z=0$ plane. The momentum 
transfer of one of the four equivalent peaks inside the first Brillouin zone is 
marked by $Q_\omega$. (c) Infinite series expansion
of $[\Gamma\chi\Gamma]^s_{\alpha\alpha';\delta\delta'}(i\nu,i\nu)$
by the Bethe-Salpeter equation. Here the polarization bubble $\chi^{0}$ is
assumed to be diagonal in orbital indices (see the main text).
(d) Same with (b) but using the frequency-averaged constant 
two-particle vertex instead of the dynamic vertex. The peak is denoted by $Q_{\rm RPA}$. 
}
\label{fig1}
\end{figure}

Figure 1(a) shows our calculated FS in the body-centered tetragonal
BZ. Two one-dimensional (1D) FS sheets ($\alpha$ and $\beta$) are almost purely 
$d_{xz}$ and $d_{yz}$-derived,
respectively, while two-dimensional (2D) FS ($\gamma$) is from $d_{xy}$, consistently with
previous studies. All local correlation effects as contained in the dynamical self energy are included.
Then we evaluated the susceptibility $\chi$, the polarization bubble $\chi^{0}$,
and the two-particle vertex $\Gamma$ which are related by the Bethe-Salpeter equation
for a given bosonic Matsubara frequency $\Omega$ {\it which is fixed to zero} in this work: 
$\chi^{s/c}_{\alpha\alpha';\beta\beta'}(i\nu,i\nu',q) = ((\chi^{0}(q))^{-1} - \Gamma^{s/c})^{-1}_
{\alpha\alpha';\beta\beta'}(i\nu,i\nu')$. All the two-particle quantities are defined in the particle-hole
channel if not explicitly indicated otherwise, and $s/c$ stands for spin/charge. 
$\Gamma^{s/c}$ is a dynamic local quantity containing all two-particle-irreducible
diagrams and is obtained in the impurity solver \cite{Hyowon,Yin2014}, as is the self energy which 
is the one-particle counterpart of the vertex. We display $\Gamma^s\chi^s\Gamma^s \equiv 
[\Gamma\chi\Gamma]^s$, which 
acts as an effective pairing potential in the Eliashberg equation, for the $d_{xy}$ orbital
channel at the lowest fermionic Matsubara frequency ($i\nu=i\nu'\approx$ 0) \cite{Matsubara} in 
Fig.~1(b). There is a strong peak at $q = (0.3,0.3) \equiv Q_\omega$ where X point is 
at $(0.5,0.5)$, reproducing the peak position of $\chi$ from measurement \cite{Steffens} 
and previous calculation \cite{Strand}. Note that $[\Gamma\chi\Gamma]^s$ has basically
the same peak structure with $\chi^s$ in BZ, since the matrix multiplication between
$\Gamma^s$ and $\chi^s$ corresponds to the weighted summation over different components of $\chi^s$ which
have similar momentum structure with one another.
To highlight the effect of frequency dependence of the vertex, we evaluate 
$[\Gamma\chi\Gamma]^s$ in Fig.~1(d) using the frequency-averaged (static) vertex corresponding 
to the equal-time component, which is equivalent to RPA. The peak is now at
$q = (0.35,0.35) \equiv Q_{\rm RPA}$ closer to X compared with $Q_\omega$ also consistently with the previous calculation \cite{Strand}, suggesting that
the momentum structure of the susceptibility alters with local correlations having different 
dynamic structures. 
$d_{xz/yz}$ components have the same peak
position with $d_{xy}$, but with a distinctively reduced amplitude,
both for dynamic and static vertices \cite{SM}
where the orbital selectivity is inscribed.
The charge counterpart, $[\Gamma\chi\Gamma]^c$, is
found to be negligible compared with $[\Gamma\chi\Gamma]^s$
for the dynamic vertex, while they are comparable 
for the static vertex \cite{SM}.
\begin{figure}[tp]
\includegraphics[width=0.8\linewidth]{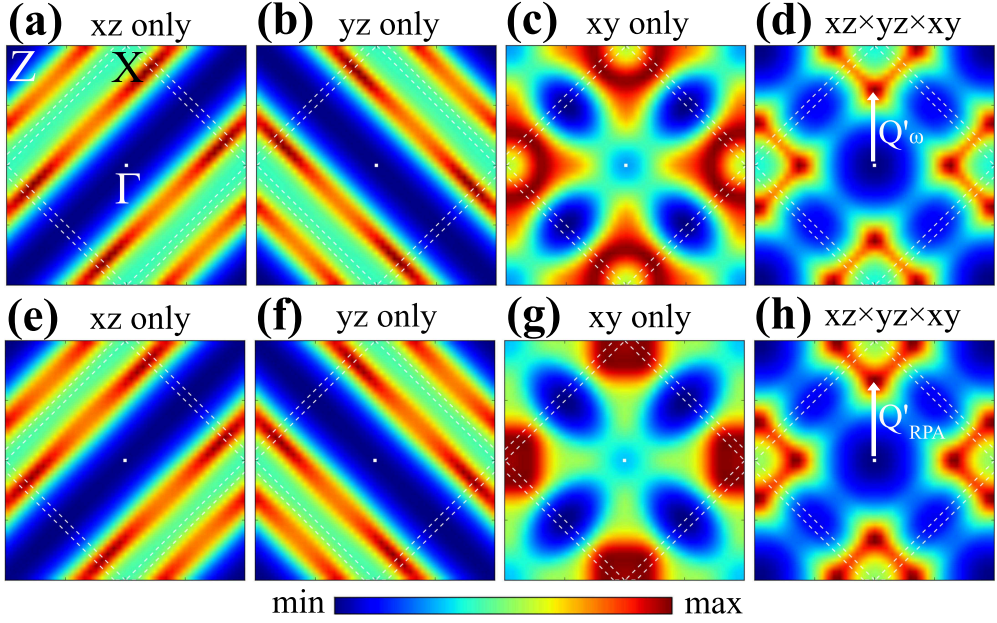}
\caption{With inter-orbital components of the vertex $\Gamma$ set to zero, $[\Gamma\chi\Gamma]^s_{\alpha\alpha;\alpha\alpha}(i\nu\approx 0,i\nu\approx 0)$ is calculated
for $\alpha=d_{xz/yz/xy}$ in (a)/(b)/(c). (e)/(f)/(g) is the same with (a)/(b)/(c), 
but using the frequency-averaged constant two-particle vertex instead of the dynamic 
vertex. (d) is obtained by multiplying (a), (b), and (c), while (h) is from multiplying
(e), (f), and (g). The peak position is denoted by $Q'_\omega$/$Q'_{\rm RPA}$ in (d)/(h).
}
\label{fig2}
\end{figure}
Now we are ready to reveal the mechanism of how the orbital selectivity encoded in the local 
vertex determines the peak position of the susceptibility in the $k$-space. To facilitate isolating
the role of each orbital in its respective dynamic structure, we calculate $[\Gamma\chi\Gamma]^s$
with inter-orbital components of the vertex turned off, i.e., 
$\Gamma^s_{\alpha\alpha';\beta\beta'}=\Gamma^s_{\alpha\alpha';\alpha\alpha'}\delta_{\alpha,\beta}\delta_{\alpha',\beta'}$, and 
display it in Fig.~2.
Since the FSs are almost completely decoupled in orbitals as seen in Fig.~1(a), the polarization bubble,
defined as the product of two counter-propagating one-particle Green's functions,
is also nearly diagonal: $\chi^{0}_{\alpha\alpha';\beta\beta'}(q) \approx \chi^{0}_{\alpha\alpha';\beta\beta'}(q)
\delta_{\alpha,\beta}\delta_{\alpha',\beta'} \equiv \chi^{0}_{\alpha\alpha'}(q)$. 
Then $[\Gamma\chi\Gamma]^s_{\alpha\alpha;\alpha\alpha}(i\nu,i\nu',q)$
is mostly composed of $\chi^{0}_{\alpha\alpha}(i\nu'',q)$ with the weight of $\Gamma^s_{\alpha\alpha;\alpha\alpha}(i\nu'',i\nu''')$
for the intermediate frequencies $i\nu''$ and $i\nu'''$, but without incorporating orbital component other than $\alpha$ in the infinite 
series as illustrated in Fig.~1(c). The diagonal component of $[\Gamma\chi\Gamma]^s$ in $d_{xz}$ orbital 
for $i\nu=i\nu'\approx 0$ is shown in Fig.~2(a). The stripe pattern along the $x$ direction is from the FS nesting in 
the $d_{xz}$-originated 1D FS sheet, and the same is true for the $d_{yz}$ component displayed in Fig.~2(b) with $\pi$/2
rotation. The fact that the $k$-dependence of $[\Gamma\chi\Gamma]^s$ is determined by the simple FS nesting 
suggests that the vertex depends only weakly on the frequency, decoupling the polarization bubble from the vertex in the frequency
domain for a given order of the infinite series in Fig. 1(c). Then the summation of the intermediate frequency produces $\sum_{\nu} \chi^{0}(i\nu) = \chi^{0}(\Omega=0)$, 
the polarization bubble in zero bosonic frequency which is characterized by the FS nesting. Indeed, 
these results are almost identical with those from frequency-averaged static vertex as displayed in Fig.~2(e) 
and Fig.~2(f).
In case of the $d_{xy}$ channel, $[\Gamma\chi\Gamma]^s$ exhibits a four-fold rotation symmetry
with weights around X points in the BZ reflecting the symmetry of the orbital as shown in both Fig.~2(c)
and Fig.~2(g) for the dynamic and frequency-averaged static vertices, respectively. On the contrary to
the $d_{xz/yz}$ channels, however, there is a noticeable difference between the two vertices.
Weights are distributed farther away from the $\Gamma$ point for the dynamic vertex, which results from the frequency-dependent coefficient
(i.e., vertex) of $\chi^{0}(i\nu)$ in the frequency ($\nu$) summation
\cite{SM}. 

As the next step, orbital-separate contributions shown in Fig.~2(a)-(c)/(e)-(g) can be
effectively coupled again to re-introduce inter-orbital components in
$[\Gamma\chi\Gamma]^s$, by simply multiplying the contributions from each 
orbital channel. This is because inter-orbital terms consisting of the product
among $\chi^0$'s  with different orbital indices, such as $\chi^0_{xy~xy}\chi^0_{xz~xz}$, are 
restored by the multiplication in the infinite series expansion of $[\Gamma\chi\Gamma]^s$ as displayed 
in Fig.~1(c).
Figure 2(d)/(h) is obtained by multiplying Fig.~2(a)/(e), (b)/(f), and (c)/(g), and indeed reproduces the strong peak structure as seen in the
original full vertex calculation result in Fig.~1(b)/(d) which includes both
intra- and inter-orbital components. Moreover, the peak position from
the dynamic vertex in Fig.~2(d) at $Q_{\omega}' \approx (0.32, 0.32)$
is closer to the zone center than that from the static vertex at
$Q_{\rm RPA}' \approx (0.35, 0.35)$, also in accordance with the
full vertex results. Having demonstrated that
$[\Gamma\chi\Gamma]^s$ can be effectively reconstructed from the product of
orbital-separate components, it is evident that the difference of the peak
position between by using the dynamic and static vertices mainly originates 
from the contrasting weight distributions of the $d_{xy}$ contribution: weight
is distributed closer to $\Gamma/X$ point in Fig.~2(c)/(g), resulting in the
peak closer to $\Gamma/X$ point in Fig.~2(d)/(h).
Therefore, we can conclude it is the $d_{xy}$ component of the vertex that is 
responsible for the shift of the peak position from $Q_{\rm RPA} = (0.35, 0.35)$,
as determined by the FS nesting, to $Q_{\omega} = (0.3, 0.3)$, via its strong frequency
dependence.
This is true not only for $\alpha=d_{xy}$ but also for $\alpha=d_{xz/yz}$ in $[\Gamma\chi\Gamma]^s_{\alpha\alpha;\alpha\alpha}$ \cite{SM} which also incorporates 
$\chi^{0}_{xy\,xy}$ through the inter-orbital vertex. The pronounced dynamic nature of the $d_{xy}$ component of the vertex is consistent with $d_{xy}$ orbital's larger mass enhancement factor pointing to the larger
correlation effect than $d_{xz/yz}$, which was suggested to originate from the proximity of the van Hove singularity 
of $d_{xy}$ band to $E_F$ \cite{MravljeAichhorn}.

Having calculated $[\Gamma\chi\Gamma]^{s/c}$, we can solve the linearized
Eliashberg equation which also has been adopted in previous works \cite{Gingras1,Gingras2,Kaser,Acharya}, derived 
from the divergence condition of the susceptibility in the particle-particle channel:
$$ -k_BT\sum_{k'\nu'\alpha'\beta'\gamma\delta}\Gamma^{pp,s/t}_{\alpha\beta;\alpha'\beta'}(k\nu;k'\nu')
\chi^{0,pp}_{\alpha'\beta'\gamma\delta}(k'\nu')\Delta_{\gamma\delta}(k'\nu')=\lambda\Delta_{\alpha
\beta}(k\nu) $$
where the eigenvalue $\lambda$ and the eigenfunction $\Delta$ can be interpreted as the pairing strength
and the gap function, respectively. $\Gamma^{pp,s/t}$ is the irreducible vertex in the particle-particle
channel and consists of different combinations of $[\Gamma\chi\Gamma]^{s/c}$ 
in interchanged momentum and orbital indices depending on spin singlet/triplet (s/t) 
pairing. Details of the formalism adopted in this study are found in Ref. \cite{Yin2014}. Here we assume the constant frequency dependence of
$\Gamma^{pp}$ so that $\Gamma^{pp}(i\nu,i\nu')=\Gamma^{pp}(i\nu\approx 0,i\nu'\approx 0)$, then the resultant gap function 
$\Delta(k)$ does not depend on the frequency either as in the BCS 
approximation where only even-frequency solutions can be captured.
Nevertheless, it should be
noted that the internal frequency summations are performed without any approximation
which is found to be essential to obtain the correct peak position of $[\Gamma\chi\Gamma]^s$ 
as demonstrated above. The gap function so obtained is distributed over entire BZ in the orbital basis
and in principle at zero energy, and hence need to be projected to the FS where spectral function 
is to be evaluated at zero energy:
$$\Delta_{ij}(k)=\sum_{\alpha\beta,i'j'}\Psi^L_{k,ii'}U_{k,i'\alpha}\Delta_{\alpha\beta}(k)
U^\dag_{k,\beta j'}\Psi^R_{k,j'j}$$. Here $\Psi^{L/R}_{k,ii'}$ is the $i$th left/right eigenvector
in $i'$th DFT band basis, of the complex DMFT eigenvalue equation at zero energy, and $U_{k,i'\alpha}$
is the projector between orbital $\alpha$ and $i'$th DFT band at $k$ \cite{Yin2014,DMFT}.
This gives us more familiar gap functions similar with those defined in the band-based mean-field 
formalism instead of Green's functions.
\begin{figure}[tp]
\includegraphics[width=0.6\linewidth]{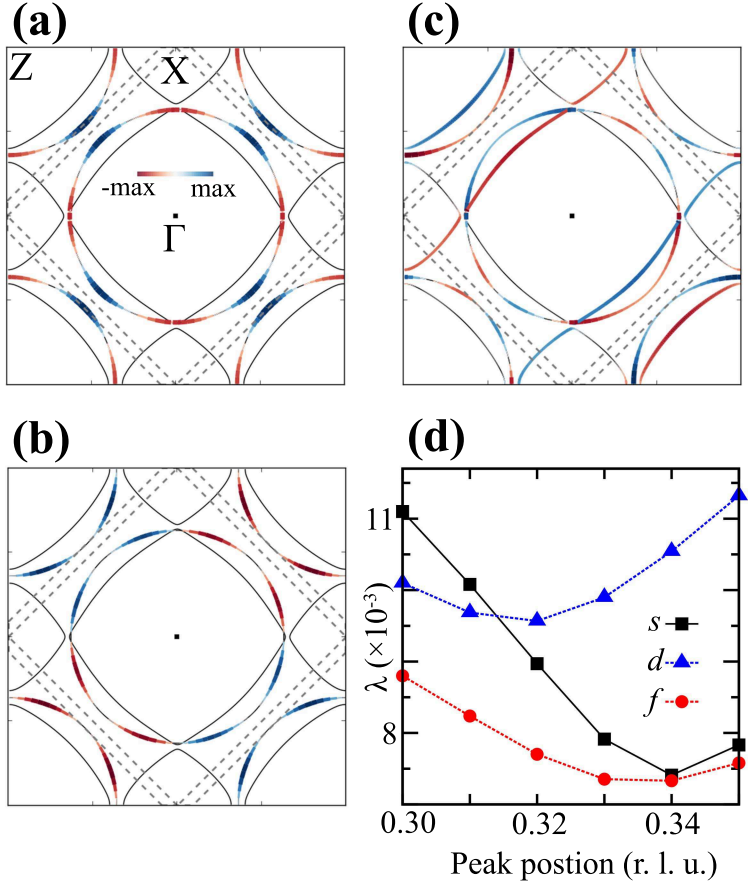}
\caption{Superconducting gap functions with three largest eigenvalues from the linearized 
Eliashberg equation, projected to the DMFT band basis on the FS, all in spin singlet. 
Gap functions in (a) and (b) are both from the $d_{xy}$ orbital with $\lambda=0.0111$ and 0.0101, 
and with $s$- and $d$-wave symmetries, respectively. $\lambda=0.0088$ solutions are doubly
degenerate, where one has the inter-orbital gap function between $d_{xz}$ and $d_{xy}$ 
as shown in (c), and the other is between $d_{yz}$ and $d_{xy}$ with the gap function 
$\pi$/2 rotated from (c). These degenerate gap functions have odd-orbital and odd-parity
$f$-wave symmetry. (d) Eigenvalue of each gap symmetry as a function of the peak
position $(q,q)$ in the spin susceptibility from $Q_\omega$ ($q$=0.30) to $Q_{\rm RPA}$ ($q$=0.35).
The peak of the spin susceptibility in an arbitrary position is approximated by
a gaussian function centered at the position. Eigenvalues are rescaled so that
they match the values from the original full calculation at $Q_\omega$.
}
\label{fig3}
\end{figure}
The gap functions for three largest eigenvalues are displayed in Fig.~3(a)-(c), all of which are in spin
singlet. The most probable solution with $\lambda=0.0111$ is from the $d_{xy}$ band and has 
a $s$-wave symmetry with nodes (''nodal s-wave'', $A_{1g}$ irreducible representation), followed by a slightly less stable solution
($\lambda=0.0101$)
with also $d_{xy}$ orbital character but in the $d_{x^2-y^2}$ symmetry ($B_{1g}$ irreducible representation). Besides the symmetry-imposed
nodes at the intersections between 2D and 1D FS sheets where the gap function changes the
sign, there are also regions with depleted weights around the van Hove point on the 2D FS
for the $d_{x^2-y^2}$ solution. The proximity of the eigenvalues or the accidental (near) degeneracy
between the $s$- and $d$-wave solutions points to the possibility of the TRSB $s+id$ order 
parameter suggested earlier \cite{RomerScherer,RomerKreisel,RomerHirschfeld,RomerMaier},
although it is $s+id_{xy}$ rather than $s+id_{x^2-y^2}$ that is more consistent with
recent experiments \cite{Benhabib,Ghosh,RomerHirschfeld}.
Meanwhile, the third
largest eigenvalue ($\lambda=0.0088$) is associated with doubly degenerate solutions which are 
protected by the lattice symmetry. Unlike $s$- and $d$-wave solutions which represent intra-orbital (band)
pairings within the $d_{xy}$ orbital, the gap functions are inter-orbital between $d_{xy}$ and $d_{xz/yz}$.
One can see that the sign of the gap function is the opposite between $d_{xy}$ and $d_{xz}$ FSs 
where they meet as in Fig.~3(c), indicating it is odd with respect to the orbital (band) interchange. 
Moreover, it is also odd in $k$-space ($f$-wave, $E_u$ irreducible representation), satisfying the antisymmetric fermion exchange
rule : $\hat{S}\hat{P}\hat{O}\hat{T}=-1\times-1\times-1\times1=-1$. Odd-parity pairing, usually
combined with spin-triplet, here associates with spin-singlet which is only possible
with the extra degree of freedom of orbitals, and hence is characteristic of the multi-orbital 
superconductivity. The degenerate pair of gap functions, one from $d_{xy}$ and $d_{xz}$ and another 
from $d_{xy}$ and $d_{yz}$ (Fig.~3(c) and its $\pi$/2 rotation, respectively), can form a TRSB 
chiral order parameter $f+if$. 
         
Our $s$-wave solution changes sign at every $\pi$/4, hence is effectively stabilized by the 
repulsive pairing potential peaked at the smaller momentum transfer $Q_\omega$, compared with 
longer $Q_{\rm RPA}$ determined by the FS nesting. $Q_\omega$ is rather short to connect the van Hove 
regions having higher density of states, hence these regions are depleted in weight for the $d$-wave
gap making it a relatively less stable solution. Indeed, Mazin {\it et al.} adopted
susceptibility with a nesting-induced peak at $Q \approx (1/3,1/3)$ to obtain a $d$-wave gap with 
maximum weights at the van Hove point in their early work \cite{Mazin}, while an $s$-wave gap has been
suggested by using the peak position determined by inelastic neutron scattering
at $Q \approx (0.3, 0.3)$ \cite{Steffens} consistently with our result. To explicitly study
the relative stability of each solution with respect to the susceptibility peak position,
we employ the spin susceptibility ($[\Gamma\chi\Gamma]^s$) consisting of 
simple gaussian peaks 
at any desired position in the BZ which should be a reasonable approximation seeing the simple peak 
structures of the original spin susceptibility shown in Fig.~1(b). 
Solving the gap equation from this susceptibility
consisting of gaussian peaks, we display the result
in Fig.~3(d). For $s$- and $d$-wave solutions, we adopt $[\Gamma\chi\Gamma]^s$  
with only $d_{xy}$
diagonal component, while for the $f$-wave gap only the inter-orbital components between $d_{xy}$
and $d_{xz/yz}$ are set non-zero. The eigenvalue $\lambda$ of each solution as a function of the
peak position is rescaled so that it retains its original value from the full calculation at
$Q_\omega$. As expected, the $s$-wave gap grows unstable with increasing momentum transfer,
while the $d$-wave gap is more stable at $Q_{\rm RPA}$. The stability of the $f$-wave gap also
decreases from $Q_\omega$ to $Q_{\rm RPA}$, so that the $d$-wave is dominant over other gap
symmetries at $Q_{\rm RPA}$. Therefore, we can conclude that TRSB $s+id$ and chiral $f+if$ gap
symmetries are stabilized by the antiferromagnetic spin fluctuation near $Q_\omega=(0.3,0.3)$,
which is only available by using the dynamical vertex function. 

Low energy excitations and the presence of (vertical line) gap nodes are another requirement
for the feasible gap symmetries imposed by experiments \cite{node1,node2,node3,node4,node5,node6}. 
Our $s$- and $d$-wave
solutions are from the $d_{xy}$ FS, so the 1D FSs from $d_{xz/yz}$ are basically gapless for
each of the two solutions (see Fig.~3(a) and (b)) and also for their TRSB complex combination 
$s+id$. When SOC is
included, the $d_{xy}$ component would be incorporated on the 1D FSs due to the orbital 
mixing, but it is likely that some part of 1D FSs are still gapless. On the other hand, 
the condition for the existence of gap nodes is tighter for the chiral $f+if$ order parameter,
since both 2D and 1D FSs participate in this inter-orbital electron pairing (Fig.~3(c) and
its $\pi$/2 rotation). Nevertheless, the van Hove points in the 2D FS would have the least weight in the 
complex combination between a sign-changing node and the relatively smaller weight
on the $\pi$/2 rotated point. Further studies might be required to elaborate on the 
compatibility of our suggested gap symmetries with experimental observations, as well as
the possible effect of SOC \cite{Gingras2,Kaser}.

In conclusion, we investigate the role of the dynamic local correlation in the spin susceptibility
and the superconducting symmetry in Sr$_2$RuO$_4$ within the DFT+DMFT framework. The two-particle vertex 
is found to weakly 
depend on the frequency within $d_{xz/yz}$ orbitals while its $d_{xy}$ component is strongly 
frequency-dependent, locating the peak of the spin susceptibility closer to the BZ center
compared with the FS nesting-driven peak position. We find that the relatively smaller momentum
transfer in the spin susceptibility stabilizes a $s$-wave gap with many sign-changing nodes
while it is against the stability of the $d$-wave, resulting in an accidental near
degeneracy of the two gap symmetries both in spin singlet. Odd-parity and odd-orbital $f$-wave
spin singlet gap functions are also found with symmetry-protected double degeneracy, which
can lead to the realization of a chiral superconductivity. Our work demonstrates how the
local correlation and orbital selectivity in a Hund's metal affect the non-local electronic 
structure in the two-particle level and the superconductivity.

\begin{acknowledgments}
The author thanks Y. Bang, I. I. Mazin, E.-G. Moon, and K. Haule for
helpful discussions.
\end{acknowledgments}



\begin{thebibliography}{99}
\bibitem{Matsubara} Because of the slight difference between the value 
at
$i\nu=i\nu'=0^\pm$ and that at $i\nu=-i\nu'=0^\pm$, where $0^{+/-}$
represents the first positive/negative fermionic Matsubara frequency, we take the
average between the two values. Same is true throughout this paper whenever this
notation appears.
\end{thebibliography}
\end{document}